\documentclass[conference,10pt]{IEEEtran}
\pdfoutput=1

\usepackage{comment}

\usepackage{cite}
\usepackage{graphicx}
\graphicspath{{images/}}

\usepackage{verbatim}

\usepackage{siunitx}

\usepackage{tikz}
\usepackage{tikzscale}
\usepackage{pgfplots}
\usepackage{float}
\usepackage{todonotes}

\usepackage[labelfont=bf]{caption}
\usepackage[caption=false]{subfig}

    \usepackage{pgfplots}
    \usepgfplotslibrary{groupplots}
    \pgfplotsset{compat=1.11,
    /pgfplots/ybar legend/.style={
    /pgfplots/legend image code/.code={%
       \draw[##1,/tikz/.cd,yshift=-0.25em]
        (0cm,0cm) rectangle (3pt,0.8em);},
   },
}

\usepackage{amssymb}
\usepackage{pifont}
\usetikzlibrary{patterns,arrows}
\usepackage{color}

\renewcommand{\IEEEbibitemsep}{0pt plus 0.5pt}
\makeatletter
\IEEEtriggercmd{\reset@font\normalfont\fontsize{7.8pt}{7.7pt}\selectfont}
\makeatother
\IEEEtriggeratref{1}

\makeatletter
\def\input@path{{images/}}
\makeatother

\ifCLASSINFOpdf
\else
\fi
\begin{document}
%
\title{Near Memory Acceleration on High Resolution Radio Astronomy Imaging}

\author{

Stefano Corda$^{1,2}$, Bram Veenboer$^{3}$, Ahsan Javed Awan$^4$, Akash Kumar$^{5}$, Roel Jordans$^1$, Henk Corporaal$^1$\\

\vspace{-0.4cm} \normalsize $^1$Eindhoven University of Technology   \hspace{0.1cm}$^2$IBM Research – Zurich  \hspace{0.1cm}$^3$Astron \hspace{0.1cm} $^4$Ericsson Research \hspace{0.1cm} $^5$TU Dresden\\\\

\{s.corda, r.jordans, h.corporaal\}@tue.nl, 
veenboer@astron.nl,
ahsan.javed.awan@ericsson.com,
akash.kumar@tu-dresden.de

}

\maketitle

\definecolor{dgreen}{rgb}{0.00, 0.75, 0.00}
\definecolor{dred}{rgb}{0.75, 0.00, 0.00}
\definecolor{dblue}{rgb}{0.00, 0.00, 0.75}
\newcommand{\stefano}[1]{[{\color{blue}Stefano: #1}]}
\newcommand{\ahsan}[1]{[{\color{red}Ahsan: #1}]}
\newcommand{\bram}[1]{[{\color{green}Bram: #1}]}

\begin{abstract}

Modern radio telescopes like the Square Kilometer Array (SKA) will need to process in real-time exabytes of radio-astronomical signals to construct a high-resolution map of the sky. Near-Memory Computing (NMC) could alleviate the performance bottlenecks due to frequent memory accesses in a state-of-the-art radio-astronomy imaging algorithm. In this paper, we show that a sub-module performing a two-dimensional fast Fourier transform (2D FFT) is memory bound using CPI breakdown analysis on IBM Power9. Then, we present an NMC approach on FPGA for 2D FFT that outperforms a CPU by up to a factor of 120x and performs comparably to a high-end GPU, while using less bandwidth and memory.

\end{abstract}


%
\IEEEpeerreviewmaketitle

\section{Introduction}
\label{sec:introduction}

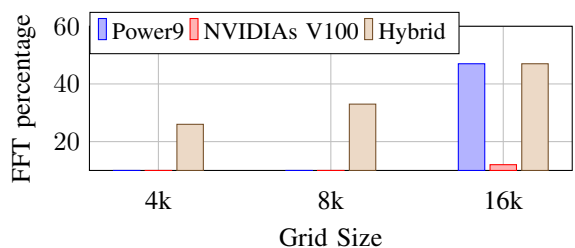
\begin{figure}[H]
  \begin{center}
  \begin{tikzpicture}
\begin{axis}[width=8cm, height=3.5cm,
legend pos=north west,
legend style={style={font=\fontsize{9}{5}\selectfont}},
legend style ={at={(0,1.1)}},
legend columns=-1,
ymin=10,
ymax=60,
xtick style={draw=none},
ytick style={draw=none},
xtick distance=1,
symbolic x coords={4k, 8k, 16k},
xlabel=Grid Size,
ylabel=FFT percentage,
ymajorgrids=true,
xmajorgrids=true,
ybar,
enlarge x limits=0.2,
]

\addplot coordinates {(4k,2)(8k,7)(16k,47)};
\addlegendentry{Power9}
\addplot coordinates {(4k,1)(8k,4)(16k,12)};
\addlegendentry{NVIDIAs V100}
\addplot coordinates {(4k,26)(8k,33)(16k,47)};
\addlegendentry{Hybrid}

\end{axis}
\end{tikzpicture}
    \caption{\emph{Percentage of the IDG execution time spent on 2D FFT. The maximum (100\%) is the total execution time of IDG. The Hybrid system consists of running the FFT on the CPU and the Gridder/Degridder on GPU. This solution has been adopted because for larger image size (e.g. 32k points per dimension) the GPU does not have sufficient memory.}
    \label{fig:motivation}}
  \end{center}
\end{figure}

The first phase of the Square Kilometre Array (SKA), the biggest radio-telescope project in the world, has very high-performance requirements, more precisely current estimates range in the order of Exaflops and PetaBytes per second \cite{van_Nieuwpoort_2010}. Especially, high-resolution image processing is crucial to detect less bright and distant sources. In particular, Image Domain Gridding (IDG) \cite{DBLP:conf/ipps/VeenboerPR17}, the state-of-the-art radio-astronomy gridding, and degridding algorithm, is a novel method employed in radio-astronomical imaging. It consists of different algorithmic steps (\emph{Fig. \ref{fig:pipeline}}), which are differently affected by image resolution.
As shown in \emph{Fig. \ref{fig:motivation}}, while some kernels, Gridder and Degridder, perform quite well even increasing the image size, we observe that the 2D FFTs exponentially become the application bottleneck since it is memory bound, and it does not reach peak performance. \emph{Fig. \ref{fig:motivation}} shows the contribution of FFT in the execution time of IDG at different image sizes. 

While the prior art has focused on accelerating gridding and degridding functions in the IDG pipeline using GPU \cite{DBLP:conf/ipps/VeenboerPR17} and FPGA \cite{DBLP:conf/europar/VeenboerR19}, we focus on accelerating the FFT function.

By employing hardware performance counters on IBM Power9, we aim to identify performance bottlenecks in the high-resolution radio-astronomy gridding and degridding imaging application. Especially memory bottlenecks could be critical in this domain like in other big-data applications~\cite{awan2017performance} due to the inefficient use of on-chip cache hierarchy, memory wall \cite{Wulf:1995:HMW:216585.216588} and the end of Dennard scaling \cite{Esmaeilzadeh:2011:DSE:2024723.2000108}. Near-Memory Computing (NMC) \cite{overviewpaper,SINGH2019102868,8806888} tries to overcome these limitations by moving the processing to where the data is located, as opposed to the classical compute-centric approach of moving the data through the entire cache hierarchy. Furthermore, NMC employs new 3D-stacked memory technologies such as HBM2 in this work, which has high off-chip memory bandwidth. We evaluate the efficacy of the NMC approach for 2D FFT acceleration. Our key contributions are:
\begin{itemize}

\item We apply a CPI breakdown analysis to the state-of-the-art radio-astronomy algorithm to detect memory bottlenecks and, we select the performance monitoring units (PMUs) on IBM Power9 that help in identifying memory boundness, in this case, represented by 2D FFTs. Furthermore, we observe in the radio-astronomy imaging context how these counters vary with the increasing image resolution and, consequently, the memory boundness.

\item We compare three different architectures showing how an NMC platform on FPGA can alleviate the memory bottleneck caused by large 2D FFTs outperforming IBM Power9 and achieving performance comparable to GPU using less memory and bandwidth.


\end{itemize}

This paper is structured as follows: \emph{Section \ref{sec:background}} presents the essential concepts on radio-astronomy and hardware performance counters. In \emph{Section \ref{sec:methodology}} we explain our methodology. Then, \emph{Section \ref{sec:results}} shows the application characterization analysis and the system evaluation. Related works are discussed in \emph{Section \ref{sec:relatedwork}} and \emph{Section \ref{sec:conclusion}} concludes the paper.

\section{Background}
\label{sec:background}

This section presents the radio-astronomy imaging (\emph{\ref{subsec:back:radio}}) and the CPI Breakdown analysis (\emph{\ref{subsec:back:cpi}}) background.

\subsection{Radio-Astronomy Imaging}
\label{subsec:back:radio}

\begin{figure}[H]
    \centering
    \includegraphics[width=9cm]{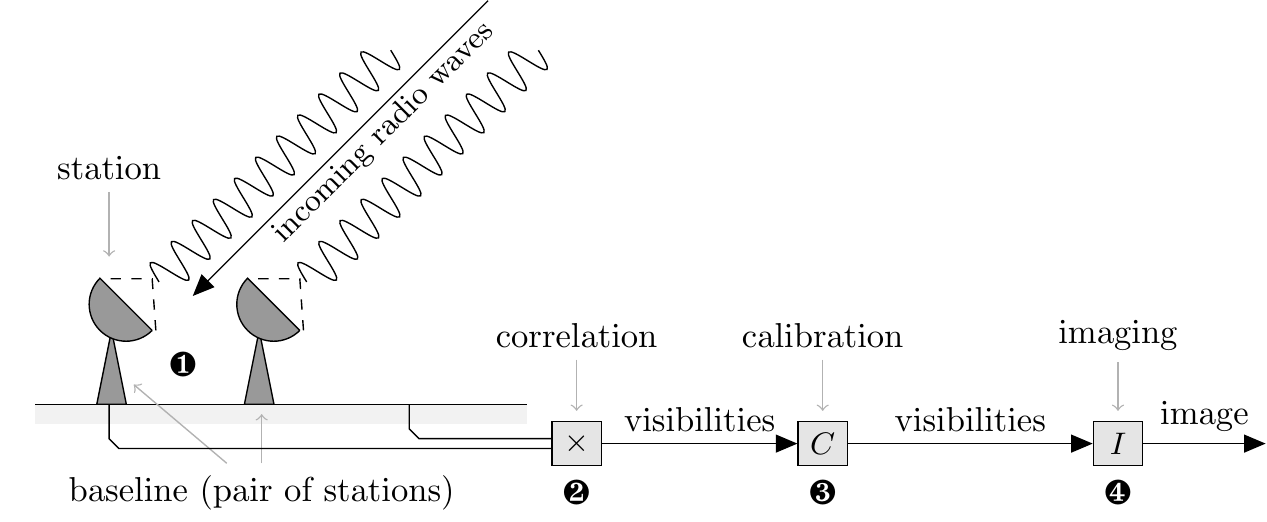}
    \caption{\emph{Radio astronomy image acquisition 
    \cite{DBLP:conf/ipps/VeenboerPR17}.}
    \label{fig:img_acq}}
\end{figure}

One of the main challenges in radio-astronomy is to translate the incoming signals from the sky to a sky image (\emph{Fig. \ref{fig:img_acq}}). This process consists of the following steps: \ding{182} digitization of the incoming electromagnetic waves from radio sources in the universe; \ding{183} correlation of the digitized signals produced by pairs of distinct stations, which produces the measurement data (\texttt{visibilities}); \ding{184} the calibration step estimates and corrects instrument parameters and environmental effects; the partially corrected visibilities are converted into a sky image by an imaging step \ding{185}.

\begin{figure}[h]
    \centering
    \includegraphics[width=9cm]{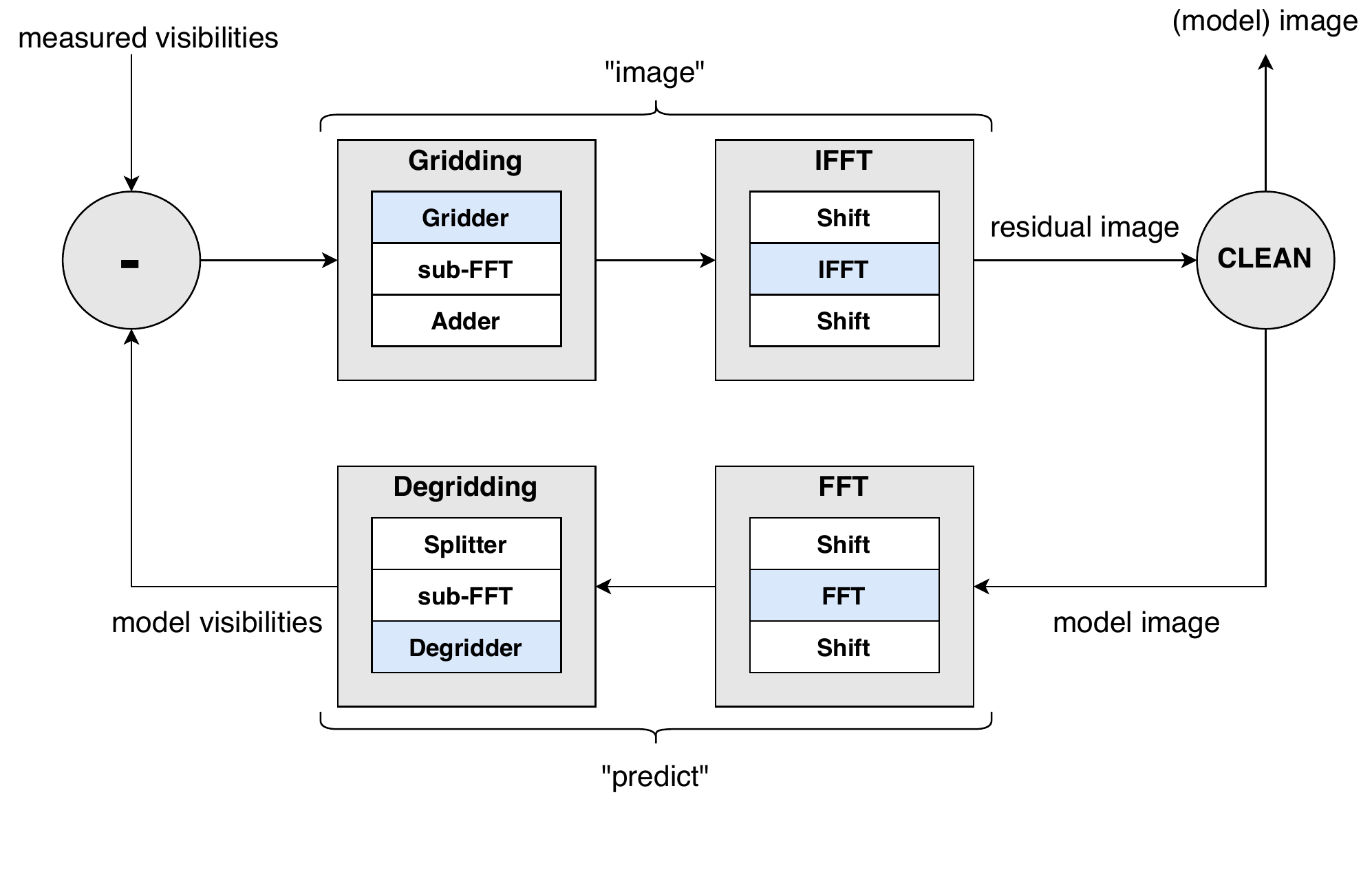}
    \caption{\emph{Complete Radio Astronomy Imaging Step \cite{DBLP:conf/ipps/VeenboerPR17}. Image Domain Gridding is a highly efficient implementation of the gridding and degridding steps (image and predict), while it leaves the CLEAN execution to other imaging application such as WSCLEAN \cite{Offringa_2014}.}
    \label{fig:pipeline}}
\end{figure}

This paper focuses on step \ding{185}, in particular on the current state-of-the-art gridding and degridding algorithm (\emph{Fig. \ref{fig:pipeline}}) called Image Domain Gridding (IDG) \cite{DBLP:conf/ipps/VeenboerPR17}.
The imaging step starts with an empty sky model and it consists of an iterative 3-steps process: \ding{172} the \texttt{imaging} step is performed on the measured visibilities producing the \texttt{residual image}; \ding{173} a variant of the CLEAN
is employed to extract one or more bright sources, which masks the more interesting weak sources, and is added to the \texttt{sky model}; \ding{174} the visibilities of the extracted sources are \texttt{predicted} and then subtracted to the input to reveal fainter sources.

Furthermore, we show in \emph{Fig. \ref{fig:pipeline}} that all these kernels consist of sub-kernels, of which the light-blue ones (Gridder, Degridder, FFT) are the ones executing most of the time (over 95\% on average for the considered image resolutions), thus focusing our work on them. 

\subsection{CPI Breakdown Analysis}
\label{subsec:back:cpi}

\begin{figure}[h]
  \begin{center}
    \includegraphics[width=9cm]{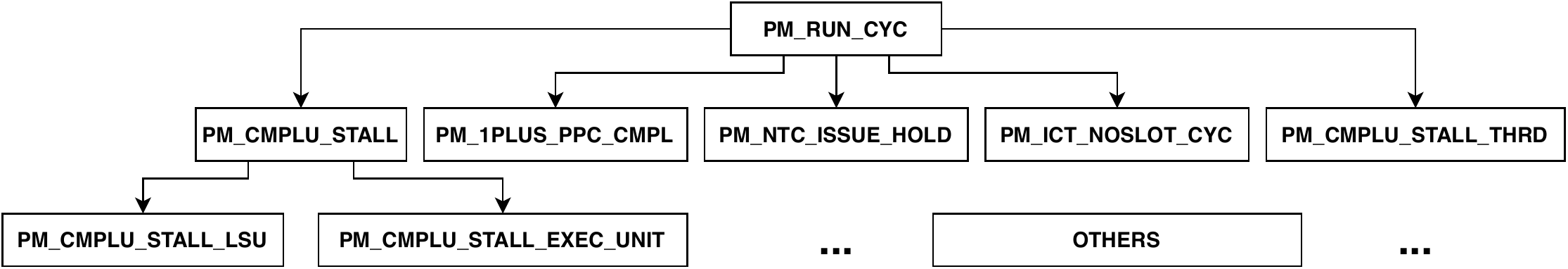}
    \caption{\emph{Power9 CPI Breakdown tree \cite{6597191}.}}
    \label{fig:cpi_tree}
  \end{center}
\end{figure}

While Intel architectures can be studied employing approaches/tools such as Top-Down \cite{6844459} or Intel VTune \cite{vtune}, IBM Power architecture lacks in this space.
In this work, we focus on the IBM Power9, which can be analyzed using the same methodology presented in \cite{6597191} for IBM Power8.

\begin{table}[h]
\centering
\caption{\emph{IBM Power9's Performance Monitoring Units (PMUs) description.} \label{tab:pmus}}
\scalebox{0.8}{
\begin{tabular}{l|l}
\textbf{PMU} & \textbf{Description} \\
\hline
 PM\_RUN\_CYC & Run cycles \\
 \hline
 PM\_CMPLU\_STALL & Nothing completed and ICT is not empty \\
 \hline
 PM\_CMPLU\_STALL\_THRD & Completion stalled because the thread was blocked \\
 \hline
 PM\_1PLUS\_PPC\_CMPL & One or more PPC instructions finished \\
  \hline
 PM\_NTC\_ISSUE\_HOLD & NTC instruction is held in the issue \\
  \hline
 PM\_ICT\_NOSLOT\_CYC & Number of cycles the ICT has no itags assigned to\\
 & this thread \\
 \hline
 PM\_CMPLU\_STALL\_LSU & Completion stalled by an LSU instruction \\
 \hline
  PM\_CMPLU\_STALL\_EXEC\_UNIT & Completion stall due to execution units\\
  & (FXU/VSU/CRU) \\
 \hline
\end{tabular}
}
\end{table}

PMUs are programmable components contained inside each microprocessor core on the chip. They are used to collect and filter information gathered from various aspects of the chip and they can attribute the events to the threads within the core. Power9 supports around 1000 PMU\footnote{https://wiki.raptorcs.com/w/images/6/6b/POWER9\_PMU\_UG\_v12\_ 28NOV2018\_pub.pdf} events that can be monitored. The CPI Breakdown consists of creating a breakdown of the total run cycles in different categories, e.g. stalls in load/store units, to understand where the application is spending most of the time, thus being able to detect application bottlenecks. A simplified representation, containing the most interesting PMUs for memory bottlenecks (see \emph{Section} \ref{sec:methodology}), of the CPI breakdown is reported in \emph{Fig. \ref{fig:cpi_tree}} and relative meaning in \emph{Tab. \ref{tab:pmus}}. 





\begin{table}[H]
\centering
\caption{\emph{System parameters and configuration.} \label{tab:systemparams}}
\scalebox{0.8}{
\begin{tabular}{l|l}
\hline
\textbf{IBM Power9 AC922} & @\SI{3.8}{GHz}, 22 cores (4-way SMT), 2 sockets,  \SI{32}{KB} L1\\
& cache per core, \SI{256}{KB} L2 cache per core, \SI{120}{MB} L3\\
& cache  per chip, \SI{512}{GB} DDR4 \SI{2666}{MHz}\\
\hline
\textbf{NVIDIA V100-SXM2-32G} & @\SI{1.53}{GHz}, 640 Tensor Cores, 5120 NVIDIA CUDA\\ & Cores,
NVlink interconnect \SI{300}{GB/s} \SI{32}{GB} HBM2 at\\ &\SI{900}{GB/s}\\
\hline
\textbf{AlphaData 9V3}            &  788 FFs, 394k LUTs, 2280 DSPs, \SI{25.3}{Mb} BRAM,\\
& \SI{90.0}{Mb} URAM, \SI{8}{GB} DDR4 \SI{2400}{MHz}\\
\hline
\textbf{AlphaData 9H7}            &  \SI{2607}{k} FFs, \SI{1304}{k} LUTs, 9024 DSPs, \SI{70.9}{Mb} BRAM,\\
& \SI{270}{Mb} URAM, \SI{8}{GB} HBM2 at \SI{460}{GB/s}\\
\hline
\end{tabular}
}

\end{table}

\section{Methodology}
\label{sec:methodology}

In this section we show the system (\emph{\ref{subsec:meth:sys}}) and the tools/software (\emph{\ref{subsec:meth:tools}}) we use for our work.

\subsection{System in use}
\label{subsec:meth:sys}

\begin{figure}[ht]
  \begin{center}
    \includegraphics[width=8cm]{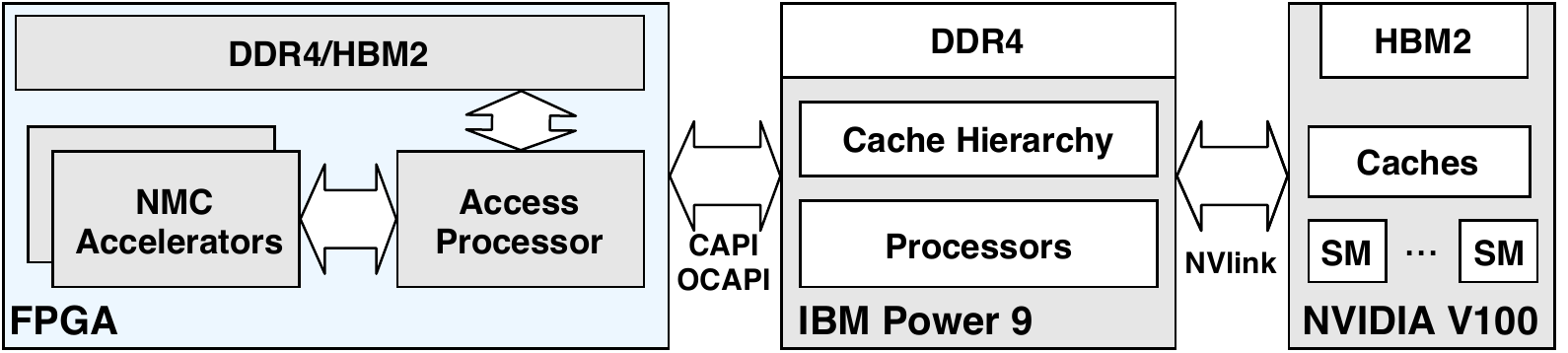}
    \caption{\emph{System employed \cite{8715088}.}}
    \label{fig:system}
  \end{center}
\end{figure}

In \emph{Fig. \ref{fig:system}}, we present the system employed for this work. We employed an IBM Power9 AC992 with 22-cores SMT4, more details are in \emph{Tab. \ref{tab:systemparams}}.
We include as a competitor to NMC an NVIDIA V100, one of the latest GPU with 32GB of HBM2 memory at \SI{900}{GB/s}, which uses similar technology to the NMC platform.
As NMC systems we use a custom hardware design called \texttt{Access Processor} (\texttt{AP}) \cite{8715088}, which can be mapped on different FPGAs (DDR4 and HBM2).

\begin{figure}[ht]
    \centering
    \includegraphics[width=9cm]{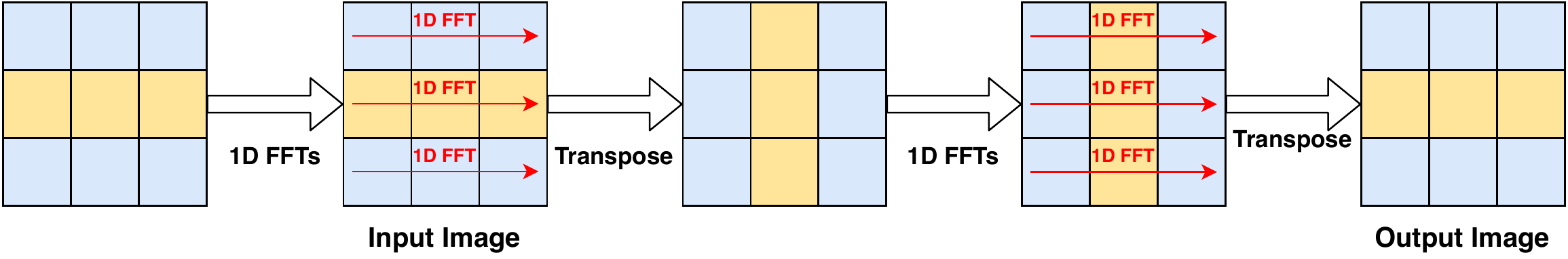}
    \caption{\emph{2D FFT decomposition in 1D FFTs and matrix transpositions.}
    \label{fig:fft_row}}
\end{figure}

Differently from a classical general-purpose computer, where the access bandwidth and latency depend on a complex mixture of workload characteristics and the memory hierarchy, the \texttt{Access Processor} (\texttt{AP}) design comprises the so-called memory controller, which has the feature of enabling more control over the memory system and programming all the concurrently running data streams from/to the attached NMC accelerators (see \emph{Fig \ref{fig:system}}). The key features of the \texttt{AP} are: 1) the B-FSM, a programmable state machine technology, applied successfully to a wide range of co-processor devices \cite{6493642}; 2) programmable address mapping scheme that can highly optimize the bandwidth utilization reducing bank conflicts and managing the data organization.
2D FFT acceleration on \texttt{AP} is performed as a combination of multiple 1D FFTs and transpose (see \emph{Fig. \ref{fig:fft_row}}). It consists of
performing a 1D FFT over all the rows of the image and an on-the-fly partial matrix transpose. Then, a 1D FFT is performed on the transposed columns of the images and they are transposed again on-the-fly. In this work we employed performance estimation, which is conservative, for the \texttt{AP} based on experiments, e.g running 1D FFTs and matrix transpose.


\subsection{Tools and Software}
\label{subsec:meth:tools}

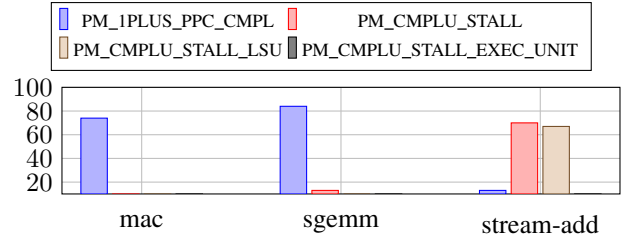
\begin{figure}[ht]
    \centering
    \begin{tikzpicture}
\begin{axis}[width=9cm, height=3cm,
legend pos=north west,
legend style={style={font=\fontsize{7}{5}\selectfont}},
legend style ={at={(-0.02,1.8)}},
legend columns=2,
ymin=10,
ymax=100,
xtick style={draw=none},
ytick style={draw=none},
xtick distance=1,
symbolic x coords={mac, sgemm, stream-add},
ymajorgrids=true,
xmajorgrids=true,
ybar,
enlarge x limits=0.2,
]

\addplot coordinates {(mac,74) (sgemm,84) (stream-add,13)};
\addlegendentry{PM\_1PLUS\_PPC\_CMPL}
\addplot coordinates {(mac,9) (sgemm,13) (stream-add,70)};
\addlegendentry{PM\_CMPLU\_STALL}
\addplot coordinates {(mac,0) (sgemm,10) (stream-add,67)};
\addlegendentry{PM\_CMPLU\_STALL\_LSU}
\addplot coordinates {(mac,9) (sgemm,3) (stream-add,2)};
\addlegendentry{PM\_CMPLU\_STALL\_EXEC\_UNIT}

\end{axis}
\end{tikzpicture}
    \caption{\emph{CPI Breakdown applied to test cases.}
    \label{fig:cpi_bench}}
\end{figure}

As a small experiment, testing our tools and analysis methodology, we show in \emph{Fig. \ref{fig:cpi_bench}} the CPI breakdown analysis (y-axis shows the PMU percentage over the total run cycles) applied to three simple benchmarks: mac, which is a compute-bound kernel written using IBM Power9 intrinsics that perform fused multiplication and accumulate over the same array of data; sgemm, which is a single-precision general matrix to matrix multiplication; stream-add, a common memory-bound benchmark used to compute the peak bandwidth of a system. We make the following two observations. First, the PMUs not included in the bar chart are nearly 0\%, thus explaining the relevance of the selected counters. Second, we can distinguish clearly a separation between a kernel completely memory bound (stream-add), which spends most of the time stalling on LSU (load-store units), and another one compute-bound (mac), which spends most of the time stalling on the computing units.

FFT was run on CPU using FFTW3 version 3.3.8 and on GPU using cuFFT of the CUDA library version 10.1. Furthermore, we improved the Degridder and Gridder algorithms on Power9 porting the Intel-based code employing IBM Power9 intrinsics. In particular, the main optimization was to use a sine/cosine lookup table, which was implemented with AltiVec intrinsics\footnote{https://openpowerfoundation.org/?resource\_lib=power-isa-version-3-0}. Especially this section of the algorithm is challenging on other CPU platforms as well; for instance on Intel high performance is obtained employing MKL (math kernel library), which is not available on PowerPC.
We use IDG\footnote{https://gitlab.com/astron-idg/idg} version \texttt{5736086c} employing the parameters in \emph{Table \ref{tab:idgparams}}.


\begin{table}[H]
\centering
\caption{\emph{Image Domain Gridding parameters.}}
    \label{tab:idgparams}
\scalebox{0.8}{
\begin{tabular}{c|c}
 \textbf{Parameters} & \textbf{Values}\\
 \hline
 Stations & 120\\
 \hline
 Channels & 16-32\\
 \hline
 Timesteps & 8192\\
 \hline
 Grid Size & 4096-8192-16384\\
 \hline
 Sub-grid Size & 32\\
 \hline
 Cycles & 1\\
 \hline
 Grid Padding & 1.0\\
 \hline
\end{tabular}
}

\end{table}

In order to characterize the application we employed \texttt{perf} \cite{perf} for extracting the PMUs values.

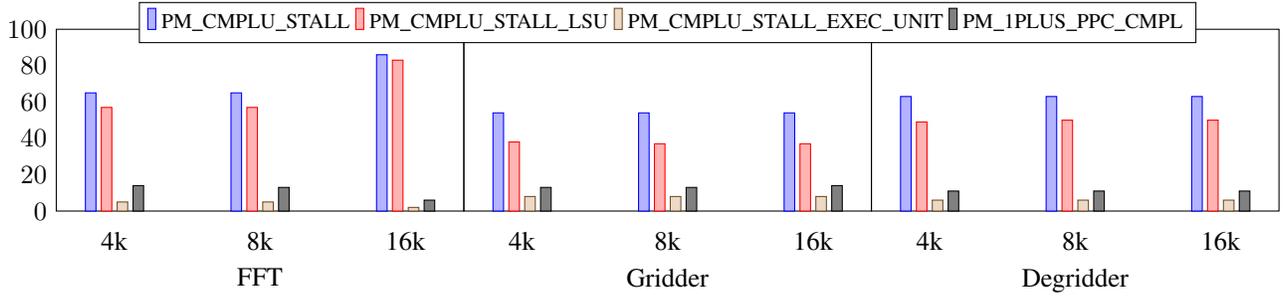
\begin{figure*}[ht]
\begin{center}
\begin{tikzpicture}
\begin{groupplot}[group style={group size=3 by 3,horizontal sep=0pt},height=4cm,width=7cm, ybar,ymin=0,ymax=100,ytick style={draw=none},xtick style={draw=none}, symbolic x coords={4k, 8k, 16k}, xtick=data]

\nextgroupplot[bar width=4pt,enlarge x limits=0.2, xlabel=FFT]
\addplot coordinates {(4k,65) (8k,65) (16k,86)};
\addplot coordinates {(4k,57) (8k,57) (16k,83)};
\addplot coordinates {(4k,5) (8k,5) (16k,2)};
\addplot coordinates {(4k,14) (8k,13) (16k,6)};

\nextgroupplot[yticklabels={},bar width=4pt,enlarge x limits=0.2,xlabel=Gridder]
\addplot coordinates {(4k,54) (8k,54) (16k,54)};
\addplot coordinates {(4k,38) (8k,37) (16k,37)};
\addplot coordinates {(4k,8) (8k,8) (16k,8)};
\addplot coordinates {(4k,13) (8k,13) (16k,14)};

\nextgroupplot[yticklabels={},bar width=4pt,enlarge x limits=0.2,legend style={at={(-0.5,1.15)},
      anchor=north,legend columns=-1,style={font=\fontsize{8}{5}\selectfont}}, xlabel=Degridder]
\addlegendentry{PM\_CMPLU\_STALL}
\addplot coordinates {(4k,63) (8k,63) (16k,63)};
\addlegendentry{PM\_CMPLU\_STALL\_LSU}
\addplot coordinates {(4k,49) (8k,50) (16k,50)};
\addlegendentry{PM\_CMPLU\_STALL\_EXEC\_UNIT}
\addplot coordinates {(4k,6) (8k,6) (16k,6)};
\addlegendentry{ PM\_1PLUS\_PPC\_CMPL}
\addplot coordinates {(4k,11) (8k,11) (16k,11)};
\end{groupplot}
\end{tikzpicture}
\end{center}
\caption{\emph{Power9 CPI Breakdown analysis of Image Domain Gridding.}\label{fig:cpi_breakdown_idg}}
\end{figure*}

\begin{figure*}[]
\centering
\subfloat[\emph{Image Domain Gridding on IBM Power9.}]{\begin{tikzpicture}
\begin{axis}[width=8.5cm, height=4cm,
legend pos=north west,
legend style={style={font=\fontsize{9}{5}\selectfont}},
legend style ={at={(0,1.1)}},
legend columns=-1,
xmax=1000,
ymin=0.001,
xmin=0.002,
ymax=50,
ymode=log, 
xmode=log,
ytick distance=10,
ymajorgrids=true,
xmajorgrids=true,
xminorgrids=true,
yminorgrids=true,
log ticks with fixed point,
xlabel=OP/Byte,
ylabel=TFLOP/s
]

\node[above,blue] at (120,1.7) {IBM Power9};

\addplot+[only marks,mark=*,mark options={scale=1.5, fill=gray},text mark as node=true, color=black] coordinates {
 (0.000001, 0.000001)
};
\addlegendentry{FFT}

\addplot+[only marks,mark=square*,mark options={scale=1.5, fill=gray},text mark as node=true, color=black] coordinates {
 (0.000001, 0.000001)
};
\addlegendentry{Gridder}

\addplot+[only marks,mark=pentagon*,mark options={scale=1.5, fill=gray},text mark as node=true, color=black] coordinates {
 (0.000001, 0.000001)
};
\addlegendentry{Degridder}

\addplot+[only marks,mark=diamond*,mark options={scale=1.5, fill=blue},text mark as node=true, color=black] coordinates {
 (0.000001, 0.000001)
};
\addlegendentry{4k}

\addplot+[only marks,mark=diamond*,mark options={scale=1.5, fill=red},text mark as node=true, color=black] coordinates {
 (0.000001, 0.000001)
};
\addlegendentry{8k}

\addplot+[only marks,mark=diamond*,mark options={scale=1.5, fill=green},text mark as node=true, color=black] coordinates {
 (0.000001, 0.000001)
};
\addlegendentry{16k}

\addplot+[only marks,mark=square*,mark options={scale=1.5, fill=blue},text mark as node=true, color=black] coordinates {
 (215.6, 0.25009)
};
\addplot+[only marks,mark=square*,mark options={scale=1.5, fill=red},text mark as node=true, color=black] coordinates {
 (214.8, 0.25562)
};
\addplot+[only marks,mark=square*,mark options={scale=1.5, fill=green},text mark as node=true, color=black] coordinates {
 (200.15, 0.25219)
};

\addplot+[only marks,mark=pentagon*,mark options={scale=1.5, fill=blue},text mark as node=true, color=black] coordinates {
 (214.7,0.25332)
};
\addplot+[only marks,mark=pentagon*,mark options={scale=1.5, fill=red},text mark as node=true, color=black] coordinates {
 (214.05,0.25258)
};
\addplot+[only marks,mark=pentagon*,mark options={scale=1.5, fill=green},text mark as node=true, color=black] coordinates {
 (199.32,0.24715)
};

\addplot+[only marks,mark=*,mark options={scale=1.5, fill=blue},text mark as node=true, color=black] coordinates {
 (6,0.02208)
};
\addplot+[only marks,mark=*,mark options={scale=1.5, fill=red},text mark as node=true, color=black] coordinates {
 (6.51,0.02603)
};
\addplot+[only marks,mark=*,mark options={scale=1.5, fill=green},text mark as node=true, color=black] coordinates {
 (7,0.00998)
};

\addplot[
    color=blue,
    ]
    coordinates {
    (0.001,0.001)(7.87,2.6752)(1000,2.6752)
    };
    
\addplot[
    color=blue,
    style=dashed,
    ]
    coordinates {
    (7.87,0.001)(7.87,2.6752)
    };    
    
\end{axis}
\end{tikzpicture}
\label{fig:roofline_idg}}
\subfloat[\emph{Image Domain Gridding on NVIDIA V100.}]{\begin{tikzpicture}
\begin{axis}[width=8.5cm, height=4cm,
legend pos=north west,
legend style={style={font=\fontsize{9}{5}\selectfont}},
legend style ={at={(0,1.1)}},
legend columns=-1,
xmax=10000,
ymin=0.001,
xmin=0.002,
ymax=1000,
ymode=log, 
xmode=log,
ytick distance=10,
ymajorgrids=true,
xmajorgrids=true,
xminorgrids=true,
yminorgrids=true,
log ticks with fixed point,
xlabel=OP/Byte,
ylabel=TFLOP/s
]

\node[above,blue] at (200,15) {NVIDIA V100};

\addplot+[only marks,mark=*,mark options={scale=1.5, fill=gray},text mark as node=true, color=black] coordinates {
 (0.000001, 0.000001)
};
\addlegendentry{FFT}

\addplot+[only marks,mark=square*,mark options={scale=1.5, fill=gray},text mark as node=true, color=black] coordinates {
 (0.000001, 0.000001)
};
\addlegendentry{Gridder}

\addplot+[only marks,mark=pentagon*,mark options={scale=1.5, fill=gray},text mark as node=true, color=black] coordinates {
 (0.000001, 0.000001)
};
\addlegendentry{Degridder}

\addplot+[only marks,mark=diamond*,mark options={scale=1.5, fill=blue},text mark as node=true, color=black] coordinates {
 (0.000001, 0.000001)
};
\addlegendentry{4k}

\addplot+[only marks,mark=diamond*,mark options={scale=1.5, fill=red},text mark as node=true, color=black] coordinates {
 (0.000001, 0.000001)
};
\addlegendentry{8k}

\addplot+[only marks,mark=diamond*,mark options={scale=1.5, fill=green},text mark as node=true, color=black] coordinates {
 (0.000001, 0.000001)
};
\addlegendentry{16k}

\addplot+[only marks,mark=square*,mark options={scale=1.5, fill=blue},text mark as node=true, color=black] coordinates {
 (702.13 , 12.315)
};
\addplot+[only marks,mark=square*,mark options={scale=1.5, fill=red},text mark as node=true, color=black] coordinates {
 (624.97 , 12.799)
};
\addplot+[only marks,mark=square*,mark options={scale=1.5, fill=green},text mark as node=true, color=black] coordinates {
 (504.98 , 12.973)
};

\addplot+[only marks,mark=pentagon*,mark options={scale=1.5, fill=blue},text mark as node=true, color=black] coordinates {
 (702.14, 11.410)
};
\addplot+[only marks,mark=pentagon*,mark options={scale=1.5, fill=red},text mark as node=true, color=black] coordinates {
 (624.97 , 11.322)
};
\addplot+[only marks,mark=pentagon*,mark options={scale=1.5, fill=green},text mark as node=true, color=black] coordinates {
 (504.85 , 11.516)
};

\addplot+[only marks,mark=*,mark options={scale=1.5, fill=blue},text mark as node=true, color=black] coordinates {
 (6, 1.335)
};
\addplot+[only marks,mark=*,mark options={scale=1.5, fill=red},text mark as node=true, color=black] coordinates {
 (6.5, 1.337)
};
\addplot+[only marks,mark=*,mark options={scale=1.5, fill=green},text mark as node=true, color=black] coordinates {
 (7, 1.180)
};

\addplot[
    color=blue,
    ]
    coordinates {
    (0.001,0.001)(17.44,15.7)(10000,15.7)
    };
    
\addplot[
    color=blue,
    style=dashed,
    ]
    coordinates {
    (17.44,0.001)(17.44,15.7)
    };    
    
\end{axis}
\end{tikzpicture}
\label{fig:roofline_nvidia}}

\hfill

\vspace{0.3cm}

\subfloat[\emph{2D FFT on \texttt{Access Processor} with DDR4.}]{\begin{tikzpicture}
\begin{axis}[width=8.5cm, height=4cm,
legend pos=north west,
legend style={style={font=\fontsize{9}{5}\selectfont}},
legend style ={at={(0,1.1)}},
legend columns=-1,
xmax=1000,
ymin=0.001,
xmin=0.002,
ymax=20,
ymode=log, 
xmode=log,
ytick distance=10,
ymajorgrids=true,
xmajorgrids=true,
xminorgrids=true,
yminorgrids=true,
log ticks with fixed point,
xlabel=OP/Byte,
ylabel=TFLOP/s
]

\node[above,blue] at (120,1) {AD9V3};

\addplot+[only marks,mark=*,mark options={scale=1.5, fill=gray},text mark as node=true, color=black] coordinates {
 (0.000001, 0.000001)
};
\addlegendentry{FFT}

\addplot+[only marks,mark=diamond*,mark options={scale=1.5, fill=blue},text mark as node=true, color=black] coordinates {
 (0.000001, 0.000001)
};
\addlegendentry{4k}

\addplot+[only marks,mark=diamond*,mark options={scale=1.5, fill=red},text mark as node=true, color=black] coordinates {
 (0.000001, 0.000001)
};
\addlegendentry{8k}

\addplot+[only marks,mark=diamond*,mark options={scale=1.5, fill=green},text mark as node=true, color=black] coordinates {
 (0.000001, 0.000001)
};
\addlegendentry{16k}

\addplot+[only marks,mark=diamond*,mark options={scale=1.5, fill=yellow},text mark as node=true, color=black] coordinates {
 (0.000001, 0.000001)
};
\addlegendentry{32k}

\addplot+[only marks,mark=*,mark options={scale=1.5, fill=blue},text mark as node=true, color=black] coordinates {
 (16.1061,0.1184274)
};
\addplot+[only marks,mark=*,mark options={scale=1.5, fill=red},text mark as node=true, color=black] coordinates {
 (17.4483,0.1302112)
};
\addplot+[only marks,mark=*,mark options={scale=1.5, fill=green},text mark as node=true, color=black] coordinates {
 (18.7905,0.1391888)
};
\addplot+[only marks,mark=*,mark options={scale=1.5, fill=yellow},text mark as node=true, color=black] coordinates {
 (20.1327,0.1464193)
};


\addplot[
    color=blue,
    ]
    coordinates {
    (0.001,0.001)(28.8,1.080)(1000,1.080)
    };
    
\addplot[
    color=blue,
    style=dashed,
    ]
    coordinates {
    (28.8,0.001)(28.8,1.080)
    };    
    
\end{axis}
\end{tikzpicture}
\label{fig:roofline_ap_ddr4}}
\subfloat[\emph{2D FFT on \texttt{Access Processor} with HBM2.}]{\begin{tikzpicture}
\begin{axis}[width=8.5cm, height=4cm,
legend pos=north west,
legend style={style={font=\fontsize{9}{5}\selectfont}},
legend style ={at={(0,1.1)}},
legend columns=-1,
xmax=1000,
ymin=0.001,
xmin=0.002,
ymax=20,
ymode=log, 
xmode=log,
ytick distance=10,
ymajorgrids=true,
xmajorgrids=true,
xminorgrids=true,
yminorgrids=true,
log ticks with fixed point,
xlabel=OP/Byte,
ylabel=TFLOP/s
]

\node[above,blue] at (120,2.5) {AD9H7};

\addplot+[only marks,mark=*,mark options={scale=1.5, fill=gray},text mark as node=true, color=black] coordinates {
 (0.000001, 0.000001)
};
\addlegendentry{FFT}

\addplot+[only marks,mark=diamond*,mark options={scale=1.5, fill=blue},text mark as node=true, color=black] coordinates {
 (0.000001, 0.000001)
};
\addlegendentry{4k}

\addplot+[only marks,mark=diamond*,mark options={scale=1.5, fill=red},text mark as node=true, color=black] coordinates {
 (0.000001, 0.000001)
};
\addlegendentry{8k}

\addplot+[only marks,mark=diamond*,mark options={scale=1.5, fill=green},text mark as node=true, color=black] coordinates {
 (0.000001, 0.000001)
};
\addlegendentry{16k}

\addplot+[only marks,mark=diamond*,mark options={scale=1.5, fill=yellow},text mark as node=true, color=black] coordinates {
 (0.000001, 0.000001)
};
\addlegendentry{32k}

\addplot+[only marks,mark=*,mark options={scale=1.5, fill=blue},text mark as node=true, color=black] coordinates {
 (16.1061,1.2583)
};
\addplot+[only marks,mark=*,mark options={scale=1.5, fill=red},text mark as node=true, color=black] coordinates {
 (17.4483,1.3848)
};
\addplot+[only marks,mark=*,mark options={scale=1.5, fill=green},text mark as node=true, color=black] coordinates {
 (18.7905,1.5032)
};
\addplot+[only marks,mark=*,mark options={scale=1.5, fill=yellow},text mark as node=true, color=black] coordinates {
 (20.1327,1.6106)
};


\addplot[
    color=blue,
    ]
    coordinates {
    (0.001,0.001)(7.9891,3.675)(1000,3.675)
    };
    
\addplot[
    color=blue,
    style=dashed,
    ]
    coordinates {
    (7.9891,0.001)(7.9891,3.675)
    };    
    
\end{axis}
\end{tikzpicture}
\label{fig:roofline_ap_hbm}}
\caption{\emph{Roofline Analysis of Image Domain Gridding.}}
\label{fig:roofline}
\end{figure*}
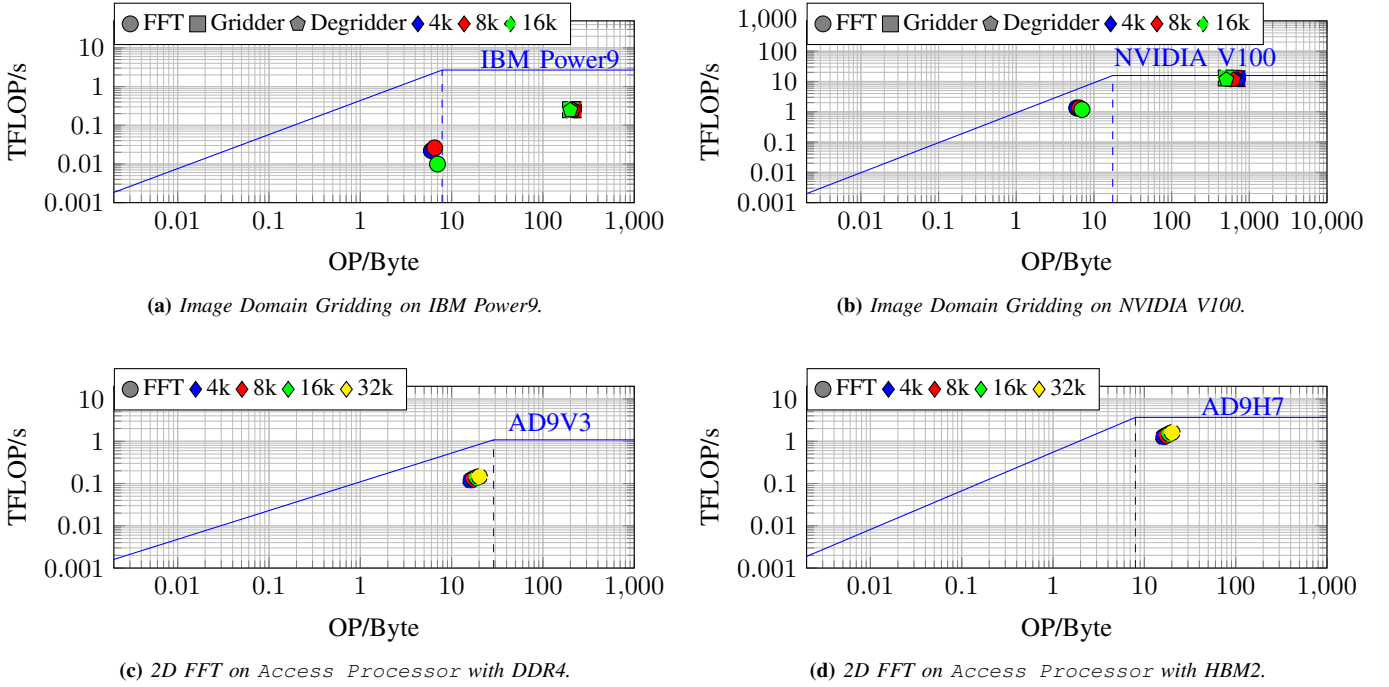

\section{Results}
\label{sec:results}

We discuss the application characterization results in \emph{\ref{subsec:res:app}} and the evaluation of three platforms for the 2D FFT kernel in \emph{\ref{subsec:res:NMC}}.

\subsection{Application Characterization}
\label{subsec:res:app}

We present the CPI breakdown analysis applied to Image Domain Gridding on IBM Power9 in \emph{Fig. \ref{fig:cpi_breakdown_idg}}. More precisely, we show the trend of the most interesting performance counters (y-axis shows the PMU percentage over the total run cycles) on increasing the visibilities grid size. The FFT spends more time on stalling on the load-store units compared to Gridder and Degridder, which means it is more memory bounded. Moreover, FFT spends less time on stalling on the execution units. Furthermore, the FFT becomes increasingly memory bound with larger grid sizes (see 16k vs 8k in \emph{Fig. \ref{fig:cpi_breakdown_idg}}) reflecting in a larger time spent on executing it (see \emph{Fig. \ref{fig:motivation}}).

We further analyze the application on IBM Power9 employing the well-known technique of the roofline model \cite{10.1145/1498765.1498785}. Power9's bandwidth is \SI{340}{GB/s} for 2 sockets and the peak performance is estimated employing the following formula:
$$TFlops=\frac{freq \; [GHz]  * \#_{op. \: per \: core} \; * \#_{cores} * \#_{sockets}}{1000}$$
Each core of the IBM Power9 can perform 16 parallel single precision operation. Using the other information from \emph{Tab. \ref{tab:systemparams}} we get \SI{2.675}{TFlops}.

\emph{Fig. \ref{fig:roofline_idg}} shows the roofline model for the kernels in Image Domain Gridding. In particular, we notice that FFT is memory bounded as it is underneath the peak bandwidth ceiling while Gridder and Degridder are compute-bound since they are underneath the peak performance ceiling. Furthermore, the FFT with a grid-size of 16k shows lower performance compared to the FFT performed with smaller grid-sizes. 
This behavior is due to the larger amount of time spent on stalling in the LSUs. Furthermore, the performance on Power9 remains low compared to the other architecture.

We also include the roofline model of Image Domain Gridding on NVIDIA V100 (see \emph{Fig. \ref{fig:roofline_nvidia}}). Peak performance is reported on the card datasheet (\SI{900}{GB/s} and \SI{15.7}{TFlops}). On NVIDIA V100 IDG achieves higher performance compared to Power9 for similar kernel characteristics. Furthermore, we build the roofline model for the 2D FFT on \texttt{Access Processor} employing the methodology proposed by Intel\footnote{https://www.intel.com/content/dam/www/programmable/us/en/pdfs/ literature/wp/wp-01222-understanding-peak-floating-point-performance-claims.pdf}. More precisely, using this methodology, which estimates the peak performance computing the maximum number of adders that can fit on the FPGA consuming all the DSPs and the logic cells, we compute the peak performance for the two FPGA boards respectively of \SI{1.080}{TFlops} and \SI{3.675}{TFlops}.
The maximum memory bandwidth is \SI{37.5}{GB/s} for 2 DDR4 bank at \SI{2400}{MHz} and \SI{460}{GB/s} for the HBM2. We show that using the FPGA with DDR4 the FFT reaches higher performance compared to Power9 and it is memory bound (see \emph{Fig. \ref{fig:roofline_ap_ddr4}}). Contrariwise, the higher bandwidth on the FPGA with HBM2 further increases the performance and makes the kernel compute-bound (see \emph{Fig. \ref{fig:roofline_ap_hbm}}). Moreover, the FPGA has similar performance compared to GPU having lower peak bandwidth and peak performance. The more efficient use of the memory is shown in \emph{Fig. \ref{fig:roofline_idg}}, where the arithmetic intensity achieved by the FPGA is higher.

\subsection{Offloading on NMC Systems}
\label{subsec:res:NMC}

\begin{figure}[]
    \centering
    \includegraphics[width=9cm]{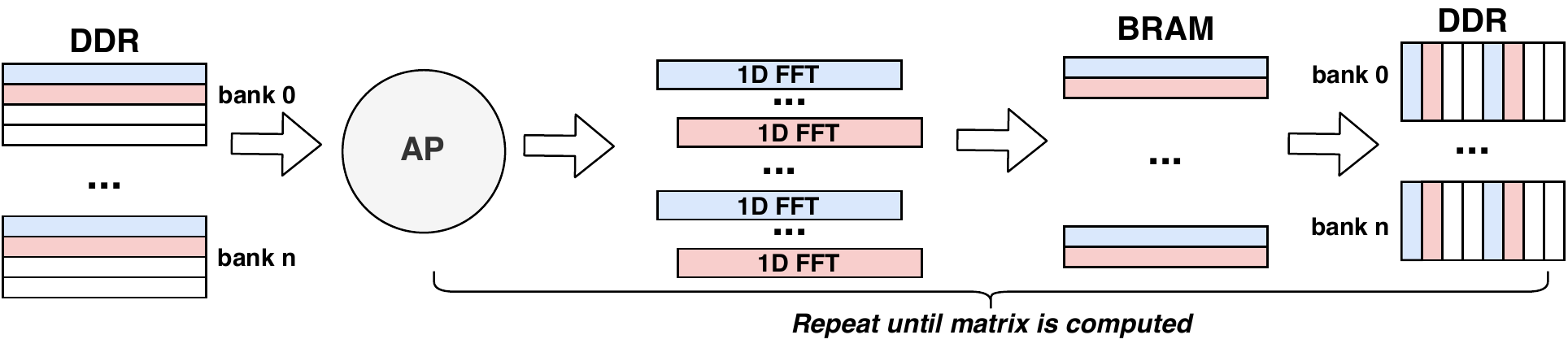}
   \caption{
   \emph{\texttt{Access Processor}'s data layout and processing for 1D FFTs over the rows showing the memory bank parallelism. This step is performed twice in order to obtain a 2D FFT.}
    \label{fig:ap_man}}
\end{figure}

The \texttt{AP} provides fine-grained control to schedule the accesses to the DDR4 and HBM2 memory (see \emph{Fig. \ref{fig:ap_man}}), the transfer of the data to and from the FPGAs internal SRAM (Block RAM and/or UltraRAM), and the processing of the data \cite{8715088}. Because the various 1D FFTs (see \emph{Fig. \ref{fig:ap_man}}) are calculated in parallel using multiple accelerators (the 1D FFTs design used is taken from \cite{7092548}), the \texttt{AP} can schedule the transfer of the input data for each 1D FFT computation from a DDR4 DIMM or HBM2 memory channel to a given accelerator during the time that additional 1D FFTs are being computed on the other accelerators. The same applies to the transfer of the 1D FFT results from an accelerator back to the DDR4 or HBM2 memory. As a result, the access, transfer, and processing of the input data and results for the 1D FFT calculations on the rows of the matrix can be overlapped in an almost seamless fashion, which enables to obtain very high performance by achieving near-optimal utilization of the available DDR4 or HBM2 memory bandwidth \cite{8686585}. In this case, the 2D FFT performance will be determined almost entirely by the available memory bandwidth, on the condition that there are enough accelerators available to fully overlap the memory access and transfer times. Experiments with FPGA cards that include DDR4\footnote{https://www.alpha-data.com/dcp/products.php?product=adm-pcie-9v3} and HBM2\footnote{https://www.alpha-data.com/dcp/products.php?product=adm-pcie-9h7} memory have been used to validate this statement.

By temporarily storing the 1D FFT results for k consecutive rows in internal memory (e.g., Block RAM), with k being equal to the number of samples fitting within the access width of the DDR4 DIMM or HBM2 memory channel (e.g., k=4 64-bit samples would fit in a 256-bit wide access vector to the HBM2 memory), the transpose can be performed on the fly when writing the k row 1D FFT results back to the DDR4 or HBM2 memory (\emph{Fig. \ref{fig:ap_man}} shows this procedure). The same operation as described above is then repeated for the columns to obtain the overall 2D FFT results over the matrix.
The effective memory access bandwidth is measured to be equal to \SI{15}{GB/s} for a single DDR4 DIMM and \SI{10}{GB/s} for a single HBM gen2 channel (which are conservative values also including the estimated impact of refresh operations, FPGA speed limitations, etc.), then the following execution times can be derived for the computation of the following four different 2D FFTs using DDR4 and HBM2 memory:

As shown in \emph{Fig. \ref{fig:motivation}} 2D FFT is the main bottleneck in IDG when enlarging the image size. We evaluate the benefits of applying NMC to FFT and comparing it to a Von-Neumann architecture. More precisely, we offload the 2D FFTs and their inverses to the \texttt{AP} design and to the NVIDIA V100. 

\begin{figure}[]
\begin{center}
\begin{tikzpicture}
\begin{axis}[width=8cm, height=3.5cm,
legend pos=north west,
legend style={style={font=\fontsize{9}{5}\selectfont}},
legend style ={at={(0,1.1)}},
legend columns=-1,
ymin=10,
ymax=100000,
ymode=log, 
xtick style={draw=none},
ytick style={draw=none},
xtick distance=1,
ytick distance=10,
symbolic x coords={AP-DDR4,AP-HBM2, V100},
ylabel=Ex. time Speedup,
ymajorgrids=true,
xmajorgrids=true,
ybar,
enlarge x limits=0.2,
]

\addplot coordinates {(AP-DDR4,429)(AP-HBM2,4558)(V100,5744)};
\addlegendentry{4k}
\addplot coordinates {(AP-DDR4,384)(AP-HBM2,4087)(V100,4933)};
\addlegendentry{8k}
\addplot coordinates {(AP-DDR4,1162)(AP-HBM2,12548)(V100,12309)};
\addlegendentry{16k}

\end{axis}
\end{tikzpicture}
\end{center}
\caption{\emph{NMC-based platform and NVIDIA V100 execution time speedup compared to IBM Power9.}\label{fig:speedup_time}}
\end{figure}
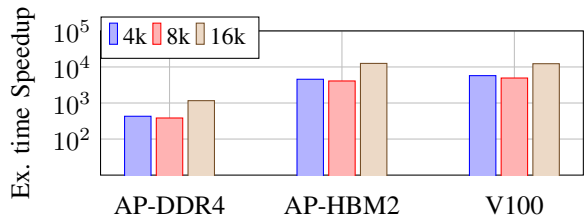

We show how a NMC approach can be faster than a common CPU (see \emph{Fig. \ref{fig:speedup_time}}) outperforming it up to 120x. The proposed design can reach similar performance compared to a high-end GPU using less memory and having a maximum bandwidth lower than half.
Furthermore, the FPGA has a lower thermal design power (TDP) compared to CPU and GPU, as reported in the data-sheets$^{1,2}$, which is 25W for the DDR4 board and 150W for the HBM2 board. Indeed, the IBM Power9 in use consumes around \SI{480}{W} when performing the FFT and the NVIDIA V100 around \SI{170}{W}. We extract the power consumption with AMESTER\footnote{https://github.com/open-power/amester} tool on the Power9 system including the NVIDIA V100. Thus, making FPGAs as good candidates for accelerating radio-astronomy applications.

\begin{table}[]
\centering
\caption{\emph{Estimated execution time of Access Processor for a single 2D FFT.}} 
    \label{tab:exap}
\scalebox{0.8}{
\begin{tabular}{c|c|c|c|c}
\hline
\textbf{Size} & \textbf{1 DDR4 DIMM} & \textbf{2 DDR4 DIMM} & \textbf{1 HBM2 channel} & \textbf{32 HBM2 channels}\\
& \textbf{\SI{15}{GB/s}} & \textbf{\SI{30}{GB/s}} & \textbf{\SI{10}{GB/s}} & \textbf{\SI{320}{GB/s}}\\
\hline
\SI{4}{k} & \SI{0.033}{s} &	\SI{0.017}{s} &	\SI{0.05}{s} &	\SI{0.0016}{s}\\
\hline
\SI{8}{k} & \SI{0.13}{s}	 & \SI{0.067}{s} & 	\SI{0.20}{s} &	\SI{0.0063}{s}\\
\hline
\SI{16}{k} & \SI{0.53}{s} &	\SI{0.27}{s} &	\SI{0.80}{s} & 	\SI{0.025}{s}\\
\hline
\SI{32}{k} & \SI{2.1}{s} &	\SI{1.1}{s} &	\SI{3.2}{s} &	\SI{0.10}{s}\\

\hline
\end{tabular}
}
\end{table}

\section{Related Work}
\label{sec:relatedwork}
In this section we provide the related work on workload characterization (\emph{\ref{subsec:rel:app}}) and on the acceleration of 2D FFT kernels (\emph{\ref{subsec:rel:fft}}).

\subsection{Application Characterization}
\label{subsec:rel:app}

A large amount of research has been focused on how to characterize workloads to detect bottlenecks. Yasin et al.~\cite{6844459} presented a similar approach to the one used in this work, but on Intel systems being the foundations of the well-known Intel VTune~\cite{vtune}. It consists of a top-down approach to identify architectural bottleneck using selected PMUs. Awan et al. use that approach to spot architectural bottlenecks in big data applications~\cite{awan2017identifying,
awan2016micro}. Differently, other approaches have been studied to characterize the application to be independent of the hardware. Corda et al. \cite{corda2019scopes,cordaDSD2019} analyzed application at LLVM-IR level to extract intrinsic application features focusing on NMC. However, PMUs are faster to be used and more accurate. As side-effects PMUs are strictly dependent on the HW employed.

\subsection{Large 2D FFT acceleration}
\label{subsec:rel:fft}

Fast Fourier Transform is one of the most widely studied algorithms in the past. Especially, large 2D FFT that is expensive on CPU, because of the enormous amount of data that must be moved from main memory through the cache hierarchy and vice-versa, have been improved.
Dang et al. \cite{Dang2014} proposed an FFT implementation on GPU clusters applied to large electromagnetic problems. 
Yu et al. \cite{Yu2011} and Akin et al.\cite{Akin2012} developed two tiling algorithms to improve performance on the 2D FFTs on different platforms.
Differently from the previous work, we employed a new computational paradigm called near-memory computing and we focused on larger 2D FFT sizes applied to radio-astronomy imaging.

\section{Conclusion}
\label{sec:conclusion}

We analyzed the state-of-the-art gridding and degridding imaging algorithm for radio-astronomy, as used in SKA, the largest radio telescope on Earth.
We employed the CPI breakdown analysis and the roofline model on IBM Power9 identifying the memory bottlenecks. Then, we showed how these bottlenecks can be alleviated by applying an NMC approach to FPGA and comparing it to CPU and GPU. Thus showing how an NMC approach can highly outperform a CPU and can achieve similar performance compared to a high-end GPU, which has higher memory bandwidth and memory size.





\section*{Acknowledgments}
This work is funded by the European Commission under Marie Sklodowska-Curie Innovative Training Networks European Industrial Doctorate (Project ID: 676240). We would like to thank Jan van Lunteren from IBM Research for providing the Access Processor and NMC accelerator architecture, and Sambit Nayak from Ericsson Research for his feedback on the draft of the paper.

\bibliographystyle{IEEEtran}
\bibliography{IEEEabrv,refshort}

\begin{thebibliography}{10}
\providecommand{\url}[1]{#1}
\csname url@samestyle\endcsname
\providecommand{\newblock}{\relax}
\providecommand{\bibinfo}[2]{#2}
\providecommand{\BIBentrySTDinterwordspacing}{\spaceskip=0pt\relax}
\providecommand{\BIBentryALTinterwordstretchfactor}{4}
\providecommand{\BIBentryALTinterwordspacing}{\spaceskip=\fontdimen2\font plus
\BIBentryALTinterwordstretchfactor\fontdimen3\font minus
  \fontdimen4\font\relax}
\providecommand{\BIBforeignlanguage}[2]{{%
\expandafter\ifx\csname l@#1\endcsname\relax
\typeout{** WARNING: IEEEtran.bst: No hyphenation pattern has been}%
\typeout{** loaded for the language `#1'. Using the pattern for}%
\typeout{** the default language instead.}%
\else
\language=\csname l@#1\endcsname
\fi
#2}}
\providecommand{\BIBdecl}{\relax}
\BIBdecl

\bibitem{van_Nieuwpoort_2010}
R.~V. van Nieuwpoort \emph{et~al.}, ``Correlating radio astronomy signals with
  many-core hardware,'' \emph{International Journal of Parallel Programming},
  vol.~39, no.~1, p. 88–114, Jun 2010.

\bibitem{DBLP:conf/ipps/VeenboerPR17}
B.~Veenboer \emph{et~al.}, ``Image-domain gridding on graphics processors,'' in
  \emph{2017 {IEEE} International Parallel and Distributed Processing
  Symposium, {IPDPS} 2017, Orlando, FL, USA, May 29 - June 2, 2017}, 2017, pp.
  545--554.

\bibitem{DBLP:conf/europar/VeenboerR19}
------, ``Radio-astronomical imaging: Fpgas vs gpus,'' in \emph{Euro-Par 2019:
  Parallel Processing - 25th International Conference on Parallel and
  Distributed Computing, G{\"{o}}ttingen, Germany, August 26-30, 2019,
  Proceedings}, 2019, pp. 509--521.

\bibitem{awan2017performance}
A.~J. Awan, ``Performance characterization and optimization of in-memory data
  analytics on a scale-up server,'' Ph.D. dissertation, KTH Royal Institute of
  Technology and Universitat Polit{\`e}cnica de Catalunya, 2017.

\bibitem{Wulf:1995:HMW:216585.216588}
W.~A. Wulf \emph{et~al.}, ``Hitting the memory wall: Implications of the
  obvious,'' \emph{SIGARCH Comput. Archit. News}, vol.~23, no.~1, pp. 20--24,
  Mar. 1995.

\bibitem{Esmaeilzadeh:2011:DSE:2024723.2000108}
H.~Esmaeilzadeh \emph{et~al.}, ``Dark silicon and the end of multicore
  scaling,'' \emph{SIGARCH Comput. Archit. News}, vol.~39, no.~3, pp. 365--376,
  Jun. 2011.

\bibitem{overviewpaper}
G.~Singh \emph{et~al.}, ``A review of near-memory computing architectures:
  Opportunities and challenges,'' in \emph{2018 21st Euromicro Conference on
  Digital System Design (DSD)}, Aug 2018, pp. 608--617.

\bibitem{SINGH2019102868}
------, ``Near-memory computing: Past, present, and future,''
  \emph{Microprocessors and Microsystems}, vol.~71, p. 102868, 2019.

\bibitem{8806888}
------, ``Napel: Near-memory computing application performance prediction via
  ensemble learning,'' in \emph{2019 56th ACM/IEEE Design Automation Conference
  (DAC)}, 2019, pp. 1--6.

\bibitem{Offringa_2014}
A.~R. Offringa \emph{et~al.}, ``Wsclean: an implementation of a fast, generic
  wide-field imager for radio astronomy,'' \emph{Monthly Notices of the Royal
  Astronomical Society}, vol. 444, no.~1, p. 606–619, Aug 2014.

\bibitem{6597191}
R.~F. {Araujo} \emph{et~al.}, ``A cpi breakdown model plug-in for optimizing
  application performance,'' in \emph{2013 3rd International Workshop on
  Developing Tools as Plug-Ins (TOPI)}, May 2013, pp. 31--36.

\bibitem{6844459}
A.~{Yasin}, ``A top-down method for performance analysis and counters
  architecture,'' in \emph{2014 IEEE International Symposium on Performance
  Analysis of Systems and Software (ISPASS)}, March 2014, pp. 35--44.

\bibitem{vtune}
\BIBentryALTinterwordspacing
INTEL. Intel vtune amplifier. [Online]. Available:
  \url{https://software.intel.com/en-us/intel-vtune-amplifier-xe}
\BIBentrySTDinterwordspacing

\bibitem{8715088}
J.~v. Lunteren \emph{et~al.}, ``Coherently attached programmable near-memory
  acceleration platform and its application to stencil processing,'' in
  \emph{2019 Design, Automation Test in Europe Conference Exhibition (DATE)},
  March 2019, pp. 668--673.

\bibitem{6493642}
------, ``Designing a programmable wire-speed regular-expression matching
  accelerator,'' in \emph{2012 45th Annual IEEE/ACM International Symposium on
  Microarchitecture}, Dec 2012, pp. 461--472.

\bibitem{perf}
\BIBentryALTinterwordspacing
Linux. Perf events tutorial. [Online]. Available:
  \url{http://perf.wiki.kernel.org/}
\BIBentrySTDinterwordspacing

\bibitem{10.1145/1498765.1498785}
S.~Williams \emph{et~al.}, ``Roofline: An insightful visual performance model
  for multicore architectures,'' \emph{Commun. ACM}, vol.~52, no.~4, p.
  65–76, Apr. 2009.

\bibitem{7092548}
H.~Giefers and et~al., ``Accelerating arithmetic kernels with coherent attached
  fpga coprocessors,'' in \emph{Proceedings of the 2015 Design, Automation \&
  Test in Europe Conference \& Exhibition}, ser. DATE ’15.\hskip 1em plus
  0.5em minus 0.4em\relax San Jose, CA, USA: EDA Consortium, 2015, p.
  1072–1077.

\bibitem{8686585}
B.~{Sukhwani} \emph{et~al.}, ``Contutto – a novel fpga-based prototyping
  platform enabling innovation in the memory subsystem of a server class
  processor,'' in \emph{2017 50th Annual IEEE/ACM International Symposium on
  Microarchitecture (MICRO)}, Oct 2017, pp. 15--26.

\bibitem{awan2017identifying}
A.~J. Awan \emph{et~al.}, ``Identifying the potential of near data processing
  for apache spark,'' in \emph{Proceedings of the International Symposium on
  Memory Systems}.\hskip 1em plus 0.5em minus 0.4em\relax ACM, 2017, pp.
  60--67.

\bibitem{awan2016micro}
------, ``Micro-architectural characterization of apache spark on batch and
  stream processing workloads,'' in \emph{2016 IEEE International Conferences
  on Big Data and Cloud Computing (BDCloud)}.\hskip 1em plus 0.5em minus
  0.4em\relax IEEE, 2016, pp. 59--66.

\bibitem{corda2019scopes}
S.~{Corda} \emph{et~al.}, ``Memory and parallelism analysis using a
  platform-independent approach,'' in \emph{ACM 22nd International Workshop on
  Software and Compilers for Embedded Systems (SCOPES '19)}.\hskip 1em plus
  0.5em minus 0.4em\relax Sankt Goar, Germany: ACM, May 2019.

\bibitem{cordaDSD2019}
------, ``Platform independent software analysis for near memory computing,''
  in \emph{2019 22nd Euromicro Conference on Digital System Design (DSD)}, Aug
  2019, pp. 606--609.

\bibitem{Dang2014}
V.~Dang and et~al., ``{GPU cluster implementation of FMM-FFT for large-scale
  electromagnetic problems},'' \emph{IEEE Antennas and Wireless Propagation
  Letters}, vol.~13, pp. 1259--1262, 2014.

\bibitem{Yu2011}
C.~L. Yu \emph{et~al.}, ``{FPGA architecture for 2d discrete Fourier transform
  based on 2d decomposition for large-sized data},'' \emph{Journal of Signal
  Processing Systems}, vol.~64, no.~1, pp. 109--122, 2011.

\bibitem{Akin2012}
B.~Akin \emph{et~al.}, ``{Memory bandwidth efficient two-dimensional fast
  Fourier transform algorithm and implementation for large problem sizes},''
  \emph{Proceedings of the 2012 IEEE 20th International Symposium on
  Field-Programmable Custom Computing Machines, FCCM 2012}, pp. 188--191, 2012.

\end{thebibliography}

\end{document}